\documentclass[preprint,12pt]{elsarticle}

\usepackage{amsmath,amssymb,amsfonts,amsthm,amscd}
\newdefinition{rmk}{Remark}

\def\div{\mathop{\mathrm{div}}\nolimits}
\def\aL{\mathsf L}
\def\aM{\mathsf M}
\def\aE{\mathsf E}
\def\aS{\mathsf S}
\def\aZ{\mathsf Z}
\def\az{\mathsf z}
\def\aF{\mathsf F}
\def\aG{\mathsf G}

\def\P{\mathcal{P}}
\def\K{\mathcal{K}}
\def\R{\mathbb{R}}
\usepackage{colonequals}
\journal{***}

\begin{document}

\begin{frontmatter}



\title{Formulation of the relativistic heat equation and the relativistic kinetic Fokker-Planck equations using GENERIC}


\author{Manh Hong Duong}

\address{Mathematics Institute,\\
University of Warwick, \\
Coventry CV4 7AL, UK. \\
Email: m.h.duong@warwick.ac.uk}

\begin{abstract}
In this paper, we formulate the relativistic heat equation and the  relativistic kinetic Fokker-Planck equations into the GENERIC (General Equation for Non-Equilibrium Reversible-Irreversible Coupling) framework. We also show that the relativistic Maxwellian distribution is the stationary solution of the latter. The GENERIC formulation provides an alternative justification that the two equations are meaningful relativistic generalizations of their non-relativistic counterparts.
\end{abstract}


\end{frontmatter}






\section{Introduction}

In this paper, we formulate the relativistic heat equation and the  relativistic kinetic Fokker-Planck equations into the GENERIC (General Equation for Non-Equilibrium Reversible-Irreversible Coupling) framework and discuss its consequences. Let us start by recalling some properties of the classical heat equation
\begin{equation}
\label{heat equation}
\partial_t\rho=\nu\Delta\rho,
\end{equation} 
and the kinetic Fokker-Planck (Kramers) equation, 
\begin{equation}
\label{KRequation}
\partial_t \rho = - \div_q\left(\rho\frac pm\right) + \div_p \left(\rho \nabla_q V\right)+ \gamma\div_p \left(\rho\frac pm\right) + \gamma \theta \Delta_p \rho.
\end{equation}
Eq. \eqref{heat equation} describes the distribution of heat (or variation in temperature) in space over time, where $\nu>0$ is the thermal diffusivity. In Eq. \eqref{KRequation}, the spatial domain is $\R^{2d}$ with coordinates $(q,p)$, with $q$ and $p$ each in $\R^d$. Subscripts as in $\div_q$ and $\div_p$ indicate that the differential operators act only on those variables. The function $V=V(q)$ is given, as are the positive constants $m,\gamma$ and $\theta$. This equation characterises the evolution of the probability density of a Brownian particle moving in a homogeneous, viscous medium and under the influence of external force fields. In this context, $q, p$ and $m$ represent the position, momentum and the mass at rest of the particle; $\gamma$ is the friction coefficient and $\theta=kT$ where $k$ is the Boltzmann constant and $T$ is the absolute temperature.

It is well-known that both Eq. \eqref{heat equation} and Eq. \eqref{KRequation} allow infinite speed of propagation. However, this violates Einstein's theory of special relativity that the speed of light $c$ is the highest admissible velocity for transport of radiation in transparent media. To remedy this issue, generalizations of Eq. \eqref{heat equation} and Eq. \eqref{KRequation} to the relativistic setting are required. For the heat equation, it was carried out first by Rosenau in \cite{Ros92} and later on by Brenier in \cite{Bre03}. Their idea is to change the classical flux based on Fourier' law, $F = −\nu\nabla \rho$, by a flux that saturates as the gradient becomes unbounded, leading to models of nonlinear degenerate parabolic equations, and in particular the following relativistic heat equation
\begin{equation}
\label{rel heat equation}
\partial_t\rho=\nu\div\left(\frac{\rho\nabla \rho}{\sqrt{\rho^2+\frac{\nu^2}{c^2}|\nabla\rho|^2}}\right),
\end{equation}
where $c$ is the speed of light.
Eq. \eqref{rel heat equation} and related models have been investigated further by many authors, see e.g., \cite{ACM06,Cas07} and references therein.

Generalization of the kinetic Fokker-Planck equation \eqref{KRequation} to the relativistic setting has been accomplished first in \cite{DMR97}. Since then various models have been considered by several authors~\cite{DH05a,DH05b,FJ07}. In contrast to the relativistic heat equation, the systems obtained in this case are still linear. In \cite{DMR97}, the authors proposed the following equation as the relativistic generalisation of \eqref{KRequation}\footnote{In \cite{DMR97}, the external potential $V\equiv 0$.}
\begin{equation}
\label{eq: rel FP DMR}
\partial_t \rho=-\div_q\left[\frac{c\, p}{\sqrt{m^2c^2+|p|^2}}\rho\right]+\div_p\left[\left(\nabla_q V+\gamma\frac{c\,p}{\sqrt{m^2c^2+|p|^2}}\right)\rho \right]+\gamma\theta\Delta_p \rho.
\end{equation}
In \cite{DH05a,DH05b} the authors suggested an alternative one, which is
\begin{align}
\label{eq: rel FP}
\partial_t \rho&=-\div_q\left[\frac{c\, p}{\sqrt{m^2c^2+|p|^2}}\rho\right]+\div_p\left[\left(\nabla_q V+\gamma\frac{p}{m}\right)\rho \right]\nonumber
\\&\qquad+\gamma\theta\div_p\left[\frac{m\,c}{\sqrt{m^2c^2+|p|^2}}\left(I+\frac{p\otimes p}{m^2c^2}\right)\nabla_p \rho\right].
\end{align}

Notations in these equations are similar to those in Eq. \eqref{KRequation}. Additionally, $I$ is the $d$-dimensional identity matrix, $p\otimes p$ denotes the $d$-dimensional matrix with entries $(p\otimes p)_{ij}=p_ip_j$, and $c$ is the speed of light. The difference between Eq. \eqref{eq: rel FP DMR} and \eqref{eq: rel FP} lies in the last two terms on the right-hand sides, namely the drift terms in $p-$ directions and the diffusion matrices. There is a large literature on these two equations, see e.g., \cite{CD07, CD08a, DC07,Dun09,Hab09a,Hab09b,Hab10,AC11,AC13} or review papers \cite{CD07a, DH09} and references therein. Many properties of them, such as $H$-theorem, fluctuation-dissipation relation and stationary solutions, have been studied in the cited papers. More particularly, in \cite{Cas07} and \cite{DH05a,DH05b,AC13} the authors have explained why \eqref{rel heat equation} and \eqref{eq: rel FP}, while having finite speed of propagation as desired, are meaningful relativistic generalizations of \eqref{heat equation} and \eqref{KRequation} respectively. The argument is that one can recover \eqref{heat equation} and \eqref{KRequation} respectively from \eqref{rel heat equation} and \eqref{eq: rel FP} as the speed of light tends to infinity, i.e., $c\rightarrow \infty$. In addition, in \cite{DH05b,AC11,AC13}, the authors suggest that Eq.~\eqref{eq: rel FP} is more physically preferable than other models derived in \cite{DH05b} because it has the relativistic Maxwellian distribution as the unique stationary solution.

The aim of this paper is to formulate the relativistic equations, \eqref{rel heat equation}, \eqref{eq: rel FP DMR} and \eqref{eq: rel FP}, into the GENERIC framework, which is a well-established framework for non-equilibrium thermodynamics. GENERIC is a formalism for non-equilibrium thermodynamics unifying both reversible and irreversible dynamics. It has been used widely in mathematical modelling of complex systems. It is now nearly impossible to list all references about GENERIC, so we will mention some recent ones. In the original papers~\cite{OG97,OG97part2}, it was originally introduced in the context of complex fluids with applications to the classical hydrodynamics and to non isothermal kinetic theory of polymeric fluid. Recently it has been applied further to anisotropic inelastic solids~\cite{HutterTervoort08a}, to viscoplastic solids~\cite{HutterTervoort08b}, to thermoelastic dissipative materials~\cite{Mielke11}, to the soft glassy rheology model~\cite{FI13}, to turbulence~\cite{Ottinger14} as well as to the damped Timoshenko and damped Bresse systems~\cite{duong2014non}. GENERIC in the relativistic setting have been investigated in, e.g., \cite{Ott98,Ott98b}. Research on GENERIC from rigorously mathematical perspectives is actively carried out, see e.g., \cite{Mielke11, DPZ13a,DPZ13b,D14b}.

Placing \eqref{rel heat equation}, \eqref{eq: rel FP DMR} and \eqref{eq: rel FP} in the GENERIC framework have a couple of benefits. Firstly, let us remind that equations \eqref{heat equation} and \eqref{KRequation} have already been cast into the GENERIC setting in \cite{ADPZ13} and \cite{DPZ13b}. We show that the GENERIC structure of \eqref{rel heat equation}, and of \eqref{eq: rel FP DMR} or \eqref{eq: rel FP} are analogous to that of \eqref{heat equation} and \eqref{KRequation} respectively, thus putting all of them in a common framework and providing an alternative justification that \eqref{rel heat equation} and \eqref{eq: rel FP DMR} or \eqref{eq: rel FP} are meaningful relativistic generalizations of \eqref{heat equation} and \eqref{KRequation} respectively. A crucial property used in the construction is a fluctuation-dissipation relation which holds true for both non-relativistic and relativistic models and is more general than the one in \cite[Eq. (13)]{DMR97}, see Remark \ref{rem: fluctuation-dissipation} for more discussion. We also show that, as a consequence of the GENERIC formulation, \eqref{eq: rel FP DMR} and \eqref{eq: rel FP} have the same stationary solution, which is the relativistic Maxwellian distribution.
It should be mentioned that stationary solutions of Eq. \eqref{eq: rel FP DMR} and Eq. \eqref{eq: rel FP} have been studied in the literature \cite{DMR97,DH05a,AC11}. Here we provide an alternative computation using the GENERIC formulation. Our approach sheds light on this issue because we show that Eq. \eqref{eq: rel FP DMR} and \eqref{eq: rel FP} actually have the same form, see \eqref{eq: rel FP2} below, and that's why they share the same stationary distribution.  Secondly, as being GENERIC systems, \eqref{rel heat equation} and \eqref{eq: rel FP} will automatically justify the first and the second law of thermodynamics. Thirdly, the GENERIC framework reveals physical and geometrical structures of \eqref{rel heat equation} and \eqref{eq: rel FP} by specifying explicitly the conservative part, the dissipative part, the energy functional and the entropy functional as well as their role as driving force of the two parts respectively. The reversible-irreversible splitting structure of the relativistic kinetic Fokker-Planck equations seems unclear in the existing papers. Last but not least, we stress that our derivation is applicable for arbitrary dimension $d$, not necessarily to separate into $1$-dimensional and $3-$ dimensional cases as in \cite{DH05a,DH05b}.

The rest of the paper is structured as follows. In Section \ref{sec: GENERIC} we summarize the GENERIC formalism in general and of \eqref{heat equation} and \eqref{KRequation} as shown in \cite{ADPZ13} and \cite{DPZ13b} respectively. We formulate the relativistic heat equation and the relativistic kinetic Fokker-Planck equations in the GENERIC framework in Section \ref{sec: GENERIC-heat} and in Section \ref{sec: GENERIC-rel} respectively. In Section \ref{sec: GENERIC-stationary}, we compute the stationary solution of \eqref{eq: rel FP}. Finally, a conclusion and discussion of further development is given in Section \ref{sec: conclusion}.

\section{GENERIC framework and its generalisation}
\label{sec: GENERIC}
\subsection{GENERIC framework}
In this section, we summarize the GENERIC framework. We refer to the original papers~\cite{OG97, OG97part2} and the book~\cite{Oettinger05} for further information. To formulate a thermodynamic system in the GENERIC framework one needs to specify five building blocks $\{\aZ,\aL,\aM,\aE,\aS\}$ and verify certain conditions put on them as following. Let $\az\in \aZ$ be a set of variables which appropriately describe the system under consideration, and $\aZ$ denotes a state space. Let $\aE, \aS\colon \aZ\rightarrow\R$ be two functionals, which are interpreted respectively as the total energy and the entropy. For each $\az\in\aZ$, let $\aL(\az)$ and $\aM(\az)$ be two operators, which are called Poisson operator and dissipative operator respectively, that map the cotangent space at $\az$ into the tangent space at $\az$. Associated with the Poisson operator and two observables, $\aF$ and $\aG$ (i.e. real
valued, sufficiently regular functionals of the set of variables
$\az$), is a Poisson bracket
\begin{equation}
\label{def:Poisson bracket}
\{\aF,\aG\}_{\aL}\colonequals \frac{\delta\aF(\az)}{\delta\az}\cdot \aL(\az)\,\frac{\delta\aG(\az)}{\delta\az},
\end{equation}
satisfying an anti-symmetric condition
\begin{equation}
\label{def: antisymmetry}
\{\aF,\aG\}_{\aL}=-\{\aG,\aF\}_{\aL},
\end{equation}
and the Jacobi identity
\begin{equation}
\label{def:Jacobi}
\{\{\aF_1,\aF_2\}_{\aL},\aF_3\}_{\aL}+\{\{\aF_2,\aF_3\}_{\aL},\aF_1\}_{\aL}+\{\{\aF_3,\aF_1\}_{\aL},\aF_2\}_{\aL}=0,
\end{equation}
for three observables $\aF_1,\aF_2,\aF_3$. Similarly associated to the dissipative operator $\aM=\aM(\az)$ and two observables, $\aF$ and $\aG$ is a dissipative bracket 
\begin{equation}
\label{def:dissipative bracket}
[\aF,\aG]_{\aM}\colonequals \frac{\delta\aF(\az)}{\delta\az}\cdot \aM(\az)\,\frac{\delta\aG(\az)}{\delta\az},
\end{equation}
which is symmetric and positive semi-definite, i.e.,
\begin{equation}
\label{def: pos def}
[\aF,\aG]_{\aM}=[\aG,\aF]_{\aM}, \quad [\aF,\aF]_{\aM}\geq 0.
\end{equation}
Moreover, $\{\aL,\aM,\aE,\aS\}$ are required to fulfill the degeneracy conditions: for all $\az\in \aZ$,
\begin{equation}
\label{def:degeneracy}
\aL(\az)\,\frac{\delta\aS(\az)}{\delta\az}=0,\quad
\aM(\az)\,\frac{\delta\aE(\az)}{\delta\az}=0.
\end{equation}
A GENERIC equation for $\az$ is then given by the following differential equation
\begin{equation}
\label{eq:GENERICeqn1}
\partial_t\az=\aL(\az)\frac{\delta\aE(\az)}{\delta\az}+\aM(\az)\frac{\delta\aS(\az)}{\delta\az},
\end{equation}
where $\partial_t\az$ is the time derivative of $\az$; $\frac{\delta\aE}{\delta\az}, \frac{\delta\aS}{\delta\az}$ are appropriate derivatives of $\aE$ and $\aS$ respectively. They could be either the Fr\'echet derivative or a gradient with respect to some inner product. Respectively, the meaning of the dot in \eqref{def:Poisson bracket} and \eqref{def:dissipative bracket} is that of the duality pairing or the formal inner product.

The degeneracy conditions~\eqref{def:degeneracy} together with the symmetries of the Poisson and dissipative brackets ensure that the energy is conserved along solutions of \eqref{eq:GENERICeqn1} and that the entropy is a non-decreasing function of time. Indeed, by straightforward calculation, we have
\begin{align*}
\frac{d\aE(\az(t))}{dt}&=\frac{\delta\aE(\az)}{\delta\az}\cdot \frac{d\az}{dt}\overset{\eqref{eq:GENERICeqn1}}{=}\frac{\delta\aE(\az)}{\delta\az}\cdot\left(\aL(\az)\frac{\delta\aE(\az)}{\delta\az}+\aM(\az)\frac{\delta\aS(\az)}{\delta\az}\right)
\\&\overset{\eqref{def:degeneracy}}{=}\frac{\delta\aE(\az)}{\delta\az}\cdot\aL(\az)\frac{\delta\aE(\az)}{\delta\az}
\overset{\eqref{def:Poisson bracket}}{=}\{\aE,\aE\}_{\aL}\overset{\eqref{def: antisymmetry}}{=}0,
\end{align*}
and
\begin{align*}
\frac{d\aS(\az(t))}{dt}&=\frac{\delta\aS(\az)}{\delta\az}\cdot \frac{d\az}{dt}\overset{\eqref{eq:GENERICeqn1}}{=}\frac{\delta\aS(\az)}{\delta\az}\cdot\left(\aL(\az)\frac{\delta\aE(\az)}{\delta\az}+\aM(\az)\frac{\delta\aS(\az)}{\delta\az}\right)
\\&\overset{\eqref{def:degeneracy}}{=}\frac{\delta\aS(\az)}{\delta\az}\cdot\aM(\az)\frac{\delta\aS(\az)}{\delta\az}\overset{\eqref{def:dissipative bracket}}=[\aS,\aS]_\aM\overset{\eqref{def: pos def}}{\geq} 0.
\end{align*}
Moreover, GENERIC allows us to determine the stationary solution. Equilibria are the maximizers of the entropy $\aS$ under the constraint that energy is constant $\aE(\az)=\aE_0$ \cite{OG97,Mielke11}. Define $\Phi(\az)=\aS(\az)-\lambda (\aE(\az)-\aE_0)$ for some $\lambda\in \R$, which plays the role of a Lagrange multiplier. The states $\az_\infty$ that maximize the entropy under the constraint $\aE(\az)=E_0$ are solutions to
\begin{equation*}
\begin{cases}
\frac{\delta \Phi(\az)}{\delta \az}=0,\\
\aE(\az)=\aE_0,
\end{cases}
\end{equation*}
which is equivalent to
\begin{equation}
\label{eq: stationary}
\begin{cases}
\frac{\delta \aS(\az)}{\delta \az}=\lambda\frac{\delta \aE(\az)}{\delta \az},\\
\aE(\az)=\aE_0.
\end{cases}
\end{equation}
If there are more constraints then the functional $\Phi$ and Eq. \eqref{eq: stationary} need to be modified accordingly. Furthermore, since $\Phi(z_t)$ is also a non-decreasing function of time, it follows that all solutions $\az_t$ of the GENERIC equation \eqref{eq:GENERICeqn1} converge to $\az_\infty$ as $t\rightarrow \infty$. We refer to \cite[Property 3]{OG97} for more detailed discussions.
\subsection{Generalised GENERIC}
 \label{sec: genralised GENERIC}
In \cite{Mielke11}, the author introduced a generalization of the GENERIC framework to the case where the dissipative part of the GENERIC evolution \eqref{eq:GENERICeqn1} does not depend linearly on gradient of the entropy $\frac{\delta \aS}{\delta \az}$. To describe the setting in \cite{Mielke11}, one needs a dissipation potential $\K$ such that for all $\az$, the function $\K(\az; \cdot)$ is convex and nonnegative and satisfies $\K(\az; 0) = 0$. The convex subdifferential of $\K(\az;\cdot)$ is defined by
\begin{equation*}
\partial_\xi \K(\az,\xi)=\{V\big| \K(\az;\tilde{\xi}) \geq \K(\az;\xi)+\langle \tilde{\xi}-\xi, V\rangle~~\text{for all}~~\tilde{\xi} \}.
\end{equation*}
If $\xi\mapsto\K(\az,\xi)$ is differentiable, then the subdifferential $\partial_\xi\K(\az,\xi)$ is the same as the derivative $\frac{\delta\K(\az,\xi)}{\delta\xi}$ of $\K(\az,\xi)$.
 
The generalized GENERIC is then given as the differential inclusion
\begin{equation}
\label{generalized GENERIC}
\partial_t\az\in \aL(\az)\frac{\delta\aE(\az)}{\delta\az}+\partial_\xi \K\Big(\az;\frac{\delta\aS(\az)}{\delta\az}\Big),
\end{equation}
and the degeneracy condition \eqref{def:degeneracy} is replaced by
\begin{equation*}
\forall \az\in \aZ,~~\xi\in \aZ^*,~~\lambda\in \R:~~ \aL(\az)\frac{\delta\aS(\az)}{\delta\az}=0,~~\text{and}~~ \K\Big(\az;\xi+\lambda\frac{\delta\aE(\az)}{\delta\az}\Big)=\K(\az;\xi).
\end{equation*}
In particular, the energy is preserved and the entropy is non-decreasing along solutions of the generalised GENERIC evolution \eqref{generalized GENERIC}. We refer to \cite[Section 2.5]{Mielke11} for more detailed discussions.
\subsection{GENERIC formulation of the heat equation and the kinetic Fokker-Planck equation}
\label{sec: GENERIC non rel}
The heat equation \eqref{heat equation} can be written as \cite{ADPZ13}
\begin{equation}
\label{GENERIC classical heat}
\partial_t \rho=\aM(\rho)\frac{\delta\aS}{\delta\rho},
\end{equation}
where 
\begin{equation}
\label{GENERIC classical heat bbl}
\aM(\rho)\xi=-\nu\div (\rho\nabla\xi),\quad \aS(\rho)=-\int \rho\log\rho,
\end{equation}
which is a special instance of the GENERIC framework, in which the conservative part is absent. Note that this GENERIC structure coincides with the Wasserstein gradient flow formulation of the heat equation \cite{JKO98,AGS08,Vil03}.

To write the kinetic Fokker-Planck equation \eqref{KRequation} in the GENERIC framework, in \cite{DPZ13b}, the authors introduced an axillary variable  $e$, depending only on $t$, so that the total energy $e(t)+\int \big(\frac{p^2}{2m}+V(q)\big)\rho_t(q,p)dqdp$ is conserved, leading to
\begin{equation}
\label{e KR}
\frac d{dt} e = \gamma \int_{\R^{2d}} \frac{p^2}{m^2}\,\rho(dqdp) - \frac{\gamma \theta d}m.
\end{equation}
The coupled system \eqref{KRequation}-\eqref{e KR} was then shown to be a GENERIC with the building blocks
\begin{equation}
\label{def:KramersInGENERIC}
\begin{aligned}
\aZ &= \P_{2,p}(\R^{2d}) \times \R,\quad & \aE(\rho,e) &= \int \Big(\frac{p^2}{2m}+V(q)\Big)\rho(q,p)dqdp + e,\\
\az& = (\rho,e),& \aS(\rho,e) &=-\theta\int \rho(q,p)\log\rho(q,p)dqdp + e,\\
\aL &= \aL(\rho,e) = \begin{pmatrix}\aL_{\rho\rho} & 0\\0 & 0\end{pmatrix},
&\quad \aM &= \aM(\rho,e) = \gamma\begin{pmatrix}\aM_{\rho\rho}  & \aM_{\rho e}\\\aM_{e\rho} & \aM_{ee}\end{pmatrix},
\end{aligned}
\end{equation}
where the operators defining $\aL$ and $\aM$ are given, upon applying them to a vector $(\xi,r)$ at $(\rho,e)$, by
\begin{alignat*}3
\aL_{\rho\rho} \xi &= \div \rho J \nabla \xi,
\\\aM_{\rho\rho}\xi &= - \div_p \rho\nabla_p \xi, &\qquad \aM_{\rho e} r &= r\div_p \Big(\rho\frac pm\Big),\\
\aM_{e\rho}\xi  &= -\int_{\R^{2d}} \frac pm \cdot\nabla_p\xi \, \rho(dqdp),
&\qquad \aM_{ee} r &= r\int_{\R^{2d}} \frac{p^2}{m^2} \,\rho(dqdp).
\end{alignat*}
Here
\begin{equation}
\label{J}
J=\begin{pmatrix}
0&-I\\
I&0
\end{pmatrix},
\end{equation}
is a $2d$- dimensional anti-symmetric matrix, and $\P_{2,p}(\R^{2d})$ is the space of probability measures on $\R^{2d}$ with bounded second $p$-moments:
\[
\P_{2,p}(\R^{2d}) := \Bigl\{ \rho\in \P(\R^{2d}): \int_{\R^{2d}} p^2 \rho(dpdq) < \infty\Bigr\}.
\]
\section{GENERIC formulation of the relativistic heat equation and of the relativistic kinetic Fokker-Planck equation}
In this section, we cast the relativistic heat equation \eqref{rel heat equation} and the relativistic kinetic Fokker-Planck equation \eqref{eq: rel FP} into the GENERIC framework and compute the stationary solution of the latter. 
\subsection{GENERIC formulation of the relativistic heat equation}
\label{sec: GENERIC-heat}
We define the dissipation potential $\K$ by
\begin{equation}
\K(\rho;\xi)=\nu\int \rho\, \varphi^*(\nabla \xi)\,dx,
\end{equation}
where
\begin{equation}
\varphi^*(z)=\frac{c^2}{\nu^2}\left(\sqrt{1+\frac{\nu^2}{c^2}|z|^2}-1\right),\quad\text{hence}	\quad \nabla \varphi^*(z)=\frac{z}{\sqrt{1+\frac{\nu^2}{c^2}|z|^2}}.
\end{equation}
It follows that
\begin{equation*}
\partial_\xi\K(\rho,\xi)=-\nu\div(\rho\nabla\varphi^*(\nabla \xi))=-\nu\div\left(\rho\frac{\nabla \xi}{\sqrt{1+\frac{\nu^2}{c^2}|\nabla\xi|^2}}\right).
\end{equation*}
Hence the relativistic heat equation \eqref{rel heat equation} can be written as
\begin{equation}
\label{GENERIC heat}
\partial_t \rho=\nu\div\left(\rho\frac{\nabla \log \rho}{\sqrt{1+\frac{\nu^2}{c^2}|\nabla \log\rho|^2}}\right)=\partial_\xi \K\Big(\rho,\frac{\delta\aS}{\delta\rho}\Big),
\end{equation}
where the entropy $\aS(\rho)$ is the negative Boltzmann entropy
\begin{equation}
\label{GEBERIC heat bbl}
 \aS(\rho)=-\int \rho\log\rho.
\end{equation}
It is straightforward to check that $\K(\rho,\cdot)$ is convex, nonnegative and satisfies $\K(\rho,0)=0$. Hence, Eq. \eqref{GENERIC heat} with the building blocks \eqref{GEBERIC heat bbl} is an instance of the generalised GENERIC evolution in which the conservative part is absent. Note that this generalised GENERIC structure coincides with the gradient flow structure of the relativistic heat equation shown in \cite{CP09}. In addition, since $\lim_{c\rightarrow \infty} \varphi^*(z)=\frac{|z|^2}{2}$, at least formally one obtains the GENERIC structure \eqref{GENERIC classical heat}-\eqref{GENERIC classical heat bbl} of the classical heat equation as $c\rightarrow \infty$. We comment on this more in Section \ref{sec: conclusion}.
\subsection{GENERIC formulation of the relativistic kinetic Fokker-Planck equation}
\label{sec: GENERIC-rel}
In this section, we show that both variants of the relativistic kinetic Fokker-Planck equation, i.e. Eq. \eqref{eq: rel FP DMR} and Eq. \eqref{eq: rel FP}, can be formulated in the GENERIC framework. We define the relativistic Hamiltonian $H$ by
\begin{equation}
H(q,p):=c\sqrt{m^2c^2+|p|^2}+V(q),
\end{equation}
and the diffusion matrix by
\begin{equation}
\mathbb{D}:=\begin{cases}
I, &\text{for Eq. \eqref{eq: rel FP DMR}},\\
D(p)=\frac{mc}{\sqrt{m^2c^2+|p|^2}}\left(I+\frac{p\otimes p}{m^2c^2}\right), &\text{for Eq. \eqref{eq: rel FP}}.
\end{cases}
\end{equation}
For each $p\in \R^{d}$, $D(p)$ is semi-positive definite since for all $\xi\in \R^d$, it holds that
\begin{align*}
D(p)\xi\cdot\xi&=\frac{1}{m(m^2c^2+|p|^2)}\sum_{i,j=1}^d(m^2c^2\delta_{ij}+p_ip_j)\xi_i\xi_j
\\&=\frac{1}{m(m^2c^2+|p|^2)}(m^2c^2\|\xi\|^2+(p\cdot \xi)^2)\geq 0.
\end{align*} 
We compute $\nabla_p H$ and $D(p)\cdot \nabla_p H$ as follows. 
We have
\begin{equation*}
\nabla_p H=\frac{c\,p}{\sqrt{m^2c^2+|p|^2}},
\end{equation*}
and
\begin{align*}
D(p)\cdot \nabla_p H&=\frac{mc}{\sqrt{m^2c^2+|p|^2}}\left(I+\frac{p\otimes p}{m^2c^2}\right)\cdot \frac{c\,p}{\sqrt{m^2c^2+|p|^2}}
\\&=\frac{1}{m(m^2c^2+|p|^2)}(m^2c^2 I+p\otimes p)\cdot p
\\&=\frac{1}{m(m^2c^2+|p|^2)}(m^2c^2 +|p|^2)\,p=\frac{p}{m}.
\end{align*}
Hence both Eq. \eqref{eq: rel FP DMR} and Eq. \eqref{eq: rel FP} can be re-written as follows
\begin{equation}
\label{eq: rel FP2}
\partial_t \rho=\div \left(J\nabla H \rho\right)+ \gamma\div_p\left[\mathbb{D}(\theta\nabla_p \rho+\rho\nabla_p H )\right],
\end{equation}
with the corresponding matrix $\mathbb{D}$, and the matrix $J$ is given by
\begin{equation*}
J=\begin{pmatrix}
0&-I\\
I&0
\end{pmatrix},
\end{equation*}
being a $2d$- dimensional anti-symmetric matrix. Let $\rho_t$ be a solution of \eqref{eq: rel FP2}.

Suppose that $V$ is non-negative and that the initial data $\rho_0$ satisfy\\ $\int_{\R^{2d}}H(q,p)\,\rho_0(dqdp)<\infty.$ We now show that $\int_{\R^{2d}}H(q,p)\,\rho_t(dqdp)$ is finite for any $t>0$. Indeed, let us compute
\begin{align}
\label{derivative of H}
&\frac{d}{dt}\int_{\R^{2d}}H(q,p)\,\rho_t(dqdp)\nonumber
\\\qquad&=\int_{\R^{2d}}H(q,p)\partial_t\rho_t(dqdp)\nonumber
\\\qquad&=\int_{\R^{2d}}H(q,p)\Big(\div \left(J\nabla H \rho_t\right)+ \gamma\div_p\left[\mathbb{D}(\theta\nabla_p \rho_t+\rho_t\nabla_p H )\right]\Big)\nonumber
\\\qquad&\overset{(*)}{=}-\gamma \int_{\R^{2d}}\mathbb{D}\nabla_pH\cdot\left[\rho_t\nabla_p H+\theta\nabla_p\rho_t\right]\,dqdp\nonumber
\\\qquad&=-\gamma \int_{\R^{2d}}\left[\mathbb{D}\nabla_p H\cdot \nabla_p H\rho_t +\theta\mathbb{D}\nabla_p H\cdot\nabla_p\rho_t\right]\,dqdp\nonumber
\\\qquad&=-\gamma\int_{\R^{2d}}\mathbb{D}\nabla_p H\cdot \nabla_p H\rho_t +\gamma\theta\int_{\R^{2d}}\div_p(\mathbb{D}\nabla_p H)\,\rho_t,
\end{align}
where in $(*)$ we have used the fact that $J$ is anti-symmetric. Note that
\begin{equation*}
\mathbb{D}\nabla_pH\cdot\nabla_p H=\begin{cases}
|\nabla_pH|^2=\frac{c^2p^2}{m^2c^2+p^2}, &\text{for Eq. \eqref{eq: rel FP DMR}}\\
\frac{c p^2}{m\sqrt{m^2c^2+p^2}},&\text{for Eq. \eqref{eq: rel FP}}.
\end{cases}
\end{equation*}
and
\begin{equation*}
\div_p(\mathbb{D}\nabla_p H)=\begin{cases}
\Delta_p H=c\Big(\frac{d}{\sqrt{m^2c^2+p^2}}-\frac{p^2}{(m^2c^2+p^2)^{\frac{3}{2}}}\Big)\leq \frac{d}{m}, &\text{for Eq. \eqref{eq: rel FP DMR}},\\
\div_p(\frac{p}{m})=\frac{d}{m},&\text{for Eq. \eqref{eq: rel FP}}.
\end{cases}
\end{equation*}
Substitute this computation to \eqref{derivative of H} we get
\begin{align}
\frac{d}{dt}\int_{\R^{2d}}H(q,p)\,\rho_t(dqdp)&=-\gamma\int_{\R^{2d}}\mathbb{D}\nabla_p H\cdot \nabla_p H\rho_t +\gamma\theta\int_{\R^{2d}}\div_p(\mathbb{D}\nabla_p H)\,\rho_t\nonumber
\\&=
\begin{cases}
-\gamma \int_{\R^{2d}}\frac{c^2p^2}{m^2c^2+p^2}\rho_t+\gamma\theta\int_{\R^{2d}}\Delta_p H\,\rho_t,&\text{for Eq. \eqref{eq: rel FP DMR}}\\
-\gamma\int_{\R^{2d}}\frac{c p^2}{m\sqrt{m^2c^2+p^2}}\,\rho_t+\frac{\gamma \theta d}{m},&\text{for Eq. \eqref{eq: rel FP}}.
\end{cases}
\end{align}

Since $\mathbb{D}\nabla_pH\cdot\nabla_p H\geq 0$, the above computation implies that 
\begin{equation}
\label{bound of derivative of H}
\frac{d}{dt}\int_{\R^{2d}}H(q,p)\,\rho_t(dqdp)\leq \frac{\gamma\theta d}{m}.
\end{equation}
Applying Gronwall's inequality, we obtain
\begin{equation}
\label{bound of H}
\int_{\R^{2d}}H(q,p)\,\rho_t(dqdp)\leq \frac{\gamma \theta d}{m} T+\int_{\R^{2d}}H(q,p)\,\rho_0(dqdp)<C,
\end{equation}
for all $t>0$ and some $C>0$. 

As a consequence, since $V\geq 0$, we have
\begin{align}
&\frac{c^2p^2}{m^2c^2+p^2}\leq c^2, \quad\frac{c p^2}{(m^2c^2+p^2)^\frac{3}{2}}\leq \frac{c\sqrt{m^2c^2+p^2}}{4m^2c^2}\leq\frac{H(q,p)}{4m^2c^2},\label{eq1}
\\&\quad\frac{c p^2}{m\sqrt{m^2c^2+p^2}}\leq \frac{c}{m}\sqrt{m^2c^2+p^2}\leq \frac{1}{m}H(q,p).\label{eq2}
\end{align}
It also follows from the above computation that $\int_{\R^{2d}} H(q,p)\rho_t(q,p)\,dqdp$ is not preserved. Similarly as in Section \ref{sec: GENERIC non rel}, we introduce an excess variable $e$, depending only on $t$, such that
\begin{equation*}
\frac{d}{dt}e(t)+\frac{d}{dt}\int_{\R^{2d}} H(q,p)\rho_t(q,p)\,dqdp =0,
\end{equation*}
which, from \eqref{derivative of H}, results in
\begin{align}
\label{eq: e}
\frac{d}{dt} e(t)&=\gamma\int_{\R^{2d}}\mathbb{D}\nabla_p H\cdot \nabla_p H\rho_t -\gamma\theta\int_{\R^{2d}}\div_p(\mathbb{D}\nabla_p H)\,\rho_t\nonumber
\\&=\begin{cases}
\gamma \int_{\R^{2d}}\frac{c^2p^2}{m^2c^2+p^2}\rho_t+c\gamma\theta\int_{\R^{2d}}\big(-\frac{d}{\sqrt{m^2c^2+p^2}}+\frac{p^2}{(m^2c^2+p^2)^{\frac{3}{2}}}\big)\,\rho_t,&\text{for Eq. \eqref{eq: rel FP DMR}}\\
\gamma\int_{\R^{2d}}\frac{c p^2}{m\sqrt{m^2c^2+p^2}}\,\rho_t-\frac{\gamma \theta d}{m},&\text{for Eq. \eqref{eq: rel FP}}.
\end{cases}
\end{align}
Note that the right-hand side of \eqref{eq: e} is well-defined due to \eqref{bound of H}, \eqref{eq1} and \eqref{eq2}.

We emphasize that the coupling is only one-direction: Eq. \eqref{eq: e} is slaved to Eq. \eqref{eq: rel FP2}. In other words, the addition of the excess variable $e$ is simply a mathematical technique, and does not change the original system. We now place the coupled system \eqref{eq: rel FP2}-\eqref{eq: e} in the GENERIC framework. The building blocks are defined as follows.
\begin{equation}
\label{def:relFPInGENERIC}
\begin{aligned}
\aZ &= \P_{2,p}(\R^{2d}) \times \R,\quad & \aE(\rho,e) &= \int_{\R^{2d}} H(q,p) \rho(q,p)\,dqdp + e,\\
\az& = (\rho,e),& \aS(\rho,e) &=
 -\theta\int_{\R^{2d}}\rho(q,p)\log\rho(q,p)\,dqdp + e,\\
\aL &= \aL(\rho,e) = \begin{pmatrix}\aL_{\rho\rho} & 0\\0 & 0\end{pmatrix},
&\quad \aM &= \aM(\rho,e) = \gamma\begin{pmatrix}\aM_{\rho\rho}  & \aM_{\rho e}\\
\aM_{e\rho} & \aM_{ee}\end{pmatrix},
\end{aligned}
\end{equation}
where the operators defining $\aL$ and $\aM$ are given, upon applying them to a vector $(\xi,r)$ at $(\rho,e)$, by
\begin{alignat*}3
\aL_{\rho\rho} \xi &= \div(\rho J \nabla \xi),\\
\aM_{\rho\rho}\xi &= - \div_p(\mathbb{D}\nabla_p \xi\,\rho), &\quad \aM_{\rho e} r &= r\div_p (\mathbb{D}\nabla_p H\,\rho),\\
\aM_{e\rho}\xi  &= -\int_{\R^{2d}}\mathbb{D}\nabla_p H  \cdot\nabla_p\xi \, \rho(dqdp),
&\quad \aM_{ee} r &= r\int_{\R^{2d}} \mathbb{D}\nabla_p H\cdot \nabla_p H\,\rho(dqdp).
\end{alignat*}

We now verify that the above building blocks indeed give rise to a GENERIC formulation of the coupled system \eqref{eq: rel FP2}-\eqref{eq: e}. First of all, we compute derivatives of $\aE$ and $\aS$
\begin{equation}
\label{eq: derivatives}
\frac{\delta\aE(\az)}{\delta\az}=\begin{pmatrix}
H(q,p)\\
1
\end{pmatrix},\quad \frac{\delta\aS(\az)}{\delta\az}=\begin{pmatrix}
-\theta(\log \rho +1)\\
1
\end{pmatrix}.
\end{equation}
Substituting these to definition of the Poisson operator and the dissipative operator, we obtain
\begin{equation*}
\aL(\az)\frac{\delta\aE(\az)}{\delta\az}=\begin{pmatrix}
\div (\rho J \nabla H)\\
0
\end{pmatrix},~~ \aM(\az)\frac{\delta\aS(\az)}{\delta\az}=\gamma\begin{pmatrix}
\div_p(\mathbb{D}\nabla_p\rho)+\div_p(\mathbb{D}\nabla_p H \rho)\\
\int_{\R^{2d}}\mathbb{D}\nabla_pH\cdot[\theta\nabla_p\rho +\nabla_pH]dqdp
\end{pmatrix}.
\end{equation*}
It follows that the equation
\begin{equation*}
\label{eq:GENERICrelFP}
\partial_t\az = \aL(\az)\frac{\delta\aE(\az)}{\delta\az}+\aM(\az)\frac{\delta\aS(\az)}{\delta\az}
\end{equation*}
is indeed the same as the coupled system \eqref{eq: rel FP2}-\eqref{eq: e}. Next, we check the conditions put on the Poisson operator $\aL$. Let $(\xi_1,r_1)$ and $(\xi_2,r_2)$ be two arbitrary vectors at $(\rho,e)$. We have 
\begin{align*}
&\langle(\xi_1,r_1),\aL(\rho,e)(\xi_2,r_2)\rangle
= \langle\xi_1,\aL_{\rho\rho}(\rho) \xi_2\rangle
= \int_{\R^{2d}} \xi_1\div\rho J\nabla \xi_2 
\\&\qquad=  \int_{\R^{2d}} \nabla \xi_1\cdot J\nabla \xi_2\,\rho=-\int_{\R^{2d}}\nabla \xi_2\cdot J\nabla \xi_1\,\rho=-\langle(\xi_2,r_2),\aL(\rho,e)(\xi_1,r_1)\rangle,
\end{align*}
where we have used that $J$ is anti-symmetric. The above equality shows that $\aL$ is antisymmetric. The verification of the Jacobi identity~\eqref{def:Jacobi} is a lengthy but elementary calculation, which  hinges on the fact that $J$ is constant and antisymmetric, and hence is omitted. Now we verify the symmetry and positive-definiteness of $\aM(\az)$. Let $(\xi_1,r_1)$ and $(\xi_2,r_2)$ be two arbitrary vectors at $(\rho,e)$, then we have
\begin{align*}
&\langle (\xi_1,r_1),\aM(\rho,e)(\xi_2,r_2)\rangle
\\&\quad=\gamma\Big[-\int\xi_1\div_p(\mathbb{D}\nabla_p\xi_2\,\rho)+r_2\int \xi_1\div_p(\mathbb{D}\nabla_p H\,\rho)-r_1\int \mathbb{D}\nabla_p H\cdot\nabla_p\xi_2\,\rho
\\&\hspace*{2cm}+r_1r_2\int \mathbb{D}\nabla_pH\cdot\nabla_p H\,\rho\Big]
\\&\quad=\gamma\Big[\int \mathbb{D}\nabla_p\xi_2\cdot\nabla_p \xi_1\,\rho-r_2\int \mathbb{D}\nabla_p H\cdot\nabla_p \xi_1\,\rho+r_1\int \xi_2\div_p(\mathbb{D}\nabla_p H\,\rho)
\\&\hspace*{2cm}+r_2r_1\int \mathbb{D}\nabla_pH\cdot\nabla_p H\,\rho\Big]
\\&\quad=\gamma\Big[-\int\xi_2\div_p(\mathbb{D}\nabla_p\xi_1\,\rho)+r_1\int \xi_2\div_p(\mathbb{D}\nabla_p H\,\rho)-r_2\int \mathbb{D}\nabla_p H\cdot\nabla_p \xi_1\,\rho
\\&\hspace*{2cm}+r_2r_1\int \mathbb{D}\nabla_pH\cdot\nabla_p H\,\rho\Big]
\\&\quad =\langle (\xi_2,r_2),\aM(\rho,e)(\xi_1,r_1)\rangle,
\end{align*}
i.e., $\aM$ is symmetric. It also follows from the above computation, by taking $(\xi_1,r_1)=(\xi_2,r_2)=(\xi,r)$, that
\begin{align*}
\langle (\xi,r),\aM(\rho,e)(\xi,r)\rangle=\gamma\int \mathbb{D}(\nabla_p\xi-r\nabla_p H)\cdot (\nabla_p\xi-r\nabla_p H)\,\rho\geq 0,
\end{align*}
i.e., $\aM$ is non-negative.

Note that the Poisson operator is the same as in Section \ref{sec: GENERIC non rel}. The coincidence of the $\aL$-operator was also recognized between the relativistic and non-relativistic imperfect fluids equations \cite{Ott98}. In addition, at least formally we have $\lim_{c\rightarrow \infty} \nabla_p H=\frac{p}{m}, \lim_{c\rightarrow \infty}D(p)= I$; hence we recover the GENERIC structure of the kinetic Fokker-Planck equation presented in Section \ref{sec: GENERIC non rel} as $c\rightarrow\infty$. We also comment on this more in Section \ref{sec: conclusion}.
\begin{rmk}
\label{rem: fluctuation-dissipation}
The crucial property, which is true for both Eq. \eqref{eq: rel FP DMR} and Eq. \eqref{eq: rel FP} and makes the calculations in this section work, is that the drift term in $p$-direction (i.e., the friction term) is exactly $\gamma\mathbb{D}\nabla H$. Note that this relation also holds true in the classical kinetic Fokker-Planck equation and is known as the fluctuation-dissipation (or Einstein) relation. It is this relation that allows us to write all these systems into a common formulation as in \eqref{eq: rel FP2}. We note that this relation is more general than the one mentioned in \cite[Eq. (13)]{DMR97} that concerns only \textit{the coefficient} of the drift term.
\end{rmk}
\subsection{Stationary distribution of the relativistic kinetic Fokker Planck equation}
\label{sec: GENERIC-stationary}
Next we seek stationary solutions of the coupled system \eqref{eq: rel FP2}-\eqref{eq: e} using \eqref{eq: stationary}. A stationary solution $\az_\infty=(\rho_\infty,e_\infty)$ of \eqref{eq: rel FP2}-\eqref{eq: e} maximizes the entropy $\aS(\az)$ under the constraints that $\aE(\az)$ is a constant and that $\int_{\R^{2d}}\rho (dqdp)=1$. Define $\Phi(\az)=\aS(\az)-\lambda_1(\aE(\az)-\aE_0)-\lambda_2 (\int_{\R^{2d}} \rho (dqdp)-1)$ for some $\lambda_1,\lambda_2\in \R$ then $(\rho_\infty,e_\infty)$ satisfies the following equation
\begin{equation*}
\begin{cases}
\frac{\delta \Phi(\az)}{\delta\az}=0,\\
\aE(\az)=\aE_0,\\
\int_{\R^{2d}}\rho(dqdp)=1.
\end{cases}
\end{equation*}
From \eqref{eq: derivatives}, this system of equations is equivalent to
\begin{equation*}
\begin{cases}
-\theta (\log \rho(q,p) +1)=\lambda_1 H(q,p)+\lambda_2,\\
1=\lambda_1,\\
\int_{\R^{2d}}H(q,p)\rho(q,p)dqdp+e=\aE_0,\\
\int_{\R^{2d}}\rho(dqdp)=1.
\end{cases}
\end{equation*}
Solving this system we obtain $\rho_\infty(dq,dp) =Z^{-1}\exp\big(-\frac{1}{\theta} H(q,p)\big)\,dqdp=Z^{-1}\exp\big(-\frac{1}{\theta} (c\sqrt{m^2c^2+|p|^2}+V(q))\big)\,dqdp$ and $e_\infty=\aE_0-\int_{\R^{2d}}H(q,p)\rho_\infty(dq,dp)$, where $Z$ is the normalizing constant
\begin{equation*}
Z=\int_{\R^{2d}}\exp\Big(-\frac{1}{\theta} (c\sqrt{m^2c^2+|p|^2}+V(q))\Big)\,dqdp.
\end{equation*}
Note that $\rho_\infty$ is exactly the relativistic Maxwellian distribution. In other words, we obtain the relativistic Maxwellian distribution as the stationary solution of the relativistic kinetic Fokker-Planck equation as a consequence of the GENERIC formulation. This stationary solution has also been obtained in \cite{DMR97,DH05b,AC11} using different arguments. Our approach works simultaneously   for both Eq. \eqref{eq: rel FP DMR} and Eq. \eqref{eq: rel FP} and interprets why they have the same stationary distribution.
\section{Conclusion and outlook}
\label{sec: conclusion}
We have unified the relativistic heat equation and the relativistic kinetic Fokker-Planck equation into the GENERIC framework  analogously to their non-relativistic counterparts. We also have established that the Maxwellian distribution is the unique stationary solution of the relativistic kinetic Fokker-Planck equation. A possibility for further
research would be to study asymptotic limits of the relativistic kinetic Fokker-Planck equation such as massless limit or the Newtonian limit, i.e., when the speed of light goes to infinity, using the GENERIC structure. It has become clear, see e.g., \cite{SandierSerfaty04,Serfaty11,AMPSV12} an a recent paper \cite{Mie14TR}, that the formulation of dissipative equations via the entropy functional and the dissipation potential is compatible well with asymptotic limits. This is because many techniques in calculus of variations, such as Gamma convergence, can be exploited. A similar framework for fully GENERIC systems has been recently proposed in \cite{DPZ13b}. We expect that these asymptotic behaviour of the relativistic equations can be obtained using the framework in the mentioned papers. We will return this issue in future publications. 
\section*{Acknowledgements}
The author would like to thank the referees for careful reading of the manuscript and their useful comments.

\end{document}